**Highlights**

   We study percolation properties of Institute-enterprise R&D Collaboration Networks.
>We derive the exact expressions for the percolation threshold of IERDCNs.
>We propose arithmetic to calculate the corresponding structural measures of IERDCNs.
>We observe the accuracy of our arithmetic, and give explanations on the discrepancies.
>We show those structural measures are useful to appraise the status of IERDCNs.

# Percolation on the institute-enterprise R&D collaboration networks

Chenguang Li,[a,*] Yongan Zhang[a]

[a] *Economics and Management School, Beijing University of Technology, Beijing 100124, China*
*Tel +8618600013270   dr.lichenguang@gmail.com*

*Abstract*: Realistic network-like systems are usually composed of multiple networks with interacting relations such as school-enterprise research and development (R&D) collaboration networks. Here we study the percolation properties of a special kind of that R&D collaboration networks, namely institute-enterprise R&D collaboration networks (IERDCNs). We introduce two actual IERDCNs to show their structural properties, and present a mathematical framework based on generating functions for analyzing an interacting network with any connection probability. Then we illustrate the percolation threshold and structural parameter arithmetic in the sub-critical and supercritical regimes. We compare the predictions of our mathematical framework and arithmetic to data for two real R&D collaboration networks and a number of simulations, and we find that they are in remarkable agreement with the data. We show applications of the framework to electronics R&D collaboration networks.

## 1. Introduction

In the past decade, complex networks have been studied intensively and widely applied in many real natural, physical and social systems. Structure and function of a single network component have already achieved great development due to numerous modeling and analyzing works [1-7]. But in fact, as one component in larger complex multiple systems, a single network does not live in isolation because it always interacts and interdepends with other networks [8]. So much attention has been focused on the topic of multiple networks with complex interplay and distinct topology recently.

Some studies on multiple networks, including interacting networks and interdependent networks are starting to demonstrate the excellent value. It is worth mentioning that several attractive models focus on properties of interdependent networks based on coupling between systems, which can be traced back to Buldyrev [9]. The purpose of these write-ups is to elucidate distinct network nodes which depend on each other and determine the robustness of networks in common [10-13]. In providing proper functionality, mutually coupled and trigger process has been emphasized that when a failure has occurred in nodes from one network, it causes nodes in the other network to fail. Furthermore, some such initial nodes' failure may trigger cascading failures from one network to another through a communication channel between a pair of nodes and even to destroy both networks [14, 15]. Beyond that, mathematical frameworks on interacting networks are another ingenious objective of multiple networks study. For instance, email, electronic commerce, electric grid, communications and socio-technical systems have been characterized by networks of networks, and the overall connectivity in these systems could be enhanced by calculating properties of components [8, 16]. Leicht and D'Souza [8] developed a framework based on generating functions for analyzing undirected interacting networks given the node connectivity within and between networks, moreover, derived exact expressions for the percolation threshold describing the onset of large-scale networks and each network individually. However, aside from that, Fu et al. [16] proposed a mathematical framework based on generating functions for analyzing directed interacting networks and derived the necessary and sufficient condition for the absence of the system-wide giant in- and out-component, and propose arithmetic to calculate the corresponding structural measures in the sub-critical and supercritical regimes. Both of their efforts extend the application of generating functions into percolation transition in multiple coupled networks.

It is generally known that regardless of individual enterprise, enterprise groups, or institutes in regional innovation systems, master cutting-edge knowledge and techniques are crucial. The R&D collaboration between institutes and enterprises is a vital form of new knowledge and technique creation. And the key element in forming R&D collaboration networks is shared knowledge and technique creation. Thus, we pay special attention to percolation on the IERDCNs because it is helpful for further study the transmission of knowledge and technique in the networks. In particular, percolation can be used to measure the number of enterprises obtain the knowledge and technique, and the giant component decides the transfer scope of the knowledge and technology.

This study is mainly focused on percolation properties of a special kind of school-enterprise R&D collaboration networks, namely Institute-enterprise R&D collaboration networks (IERDCNs). Furthermore, we define R&D agents as nodes, and collaborations as edges. There are two types of nodes as different networks in IERDCNs, one is technology enterprises (hereinafter referred to as enterprises), and the other is research institutes (hereinafter referred to as institutes, which include colleges and private research institutes). They always show their independent in the respective network, because of the intense competition between enterprises in the same industry. If one partner breaks down, others will still work. So IERDCNs are interacting networks containing connectivity links only, and it is appropriate to choose the mathematical framework created by Leicht and D'Souza [8] as the base model. Something interesting and distinct from prior studies are we tried to take connection probability into consideration. Moreover, conditions for these components to become the giant ones are worthy to be discussed. Hence, further investigations are needed to model the mechanism underlying discontinuous percolation processes. Our work can supplement and enrich existing studies in multiple networks.

The rest of the paper is organized as follows. From degree distribution, density, and assortativity, etc., Section 2 introduces the unique network topology and structural properties of IERDCNs. Section 3 puts forward two mathematical frameworks for IERDCNs, a general one and a special one with connection probability, which is useful for deriving percolation conditions and calculating the average sizes of components. We evaluate our arithmetic using a set of simulation instances and discuss the practical application of IERDCNs in Section 4. Finally, in Section 5, we discuss possible implications and extensions.

## 2. Structural properties

As mentioned previously, enterprises and institutes compose the IERDCNs which can be found mostly in technology innovation networks [17]. Resource-based theory emphasizes that there is the heterogeneous resource in the enterprise and institutes, in which the percolation could be able to create the synergy effect of knowledge and technique flow in overall R&D collaboration systems. Meanwhile, any enterprise may not satisfy the need for innovation by utilizing its internal limited resource. Instead, it needs R&D collaboration to get more resources, knowledge and techniques. Moreover, enterprises in the same industry tend to select institutes as partners because of intense inter-firm competition caused by homogeneous commodity or services. Institutes with abundant intellectual resource have a team of professional researchers and technicians. By external collaboration, institutes may transmit their accumulated knowledge and generate the intellectual property rights of new technology. Certainly, they can obtain economic benefits. For instance, ENEA is an Italian Government sponsored R&D center. If an enterprise has innovative ideas and wants to invent a technique, it will entrust one or two institutes of ENEA with the development of cutting-edge knowledge and technique, after all research activities over, they will apply for a patent and share its ownership. Thus, enterprises can greatly enhance their current workflow, productivity and quality with the techniques, and institutes can increase their efforts on promoting new technology when collaborate with other enterprises.

We have found some interesting things that most enterprises tend to select two institutes for R&D collaboration. That may be a necessary safeguard, in order to prevent the failure of single connection. As well as there are few collaborations of enterprise to enterprise for inter-firm competition we mentioned. Furthermore, there are also few of internal connections between institutes. The cause may be each institute has the abundant intellectual resources.

As already stated in our letter, an IERDCN is comprised of enterprise R&D network and institute R&D network. In enterprise R&D network, different enterprises have their partner selection preferences, but enterprises in the same industry have their special uniform characteristics on collaborative R&D, hence here we take the electronics enterprises into this study and investigate their combinations with the institute R&D network.

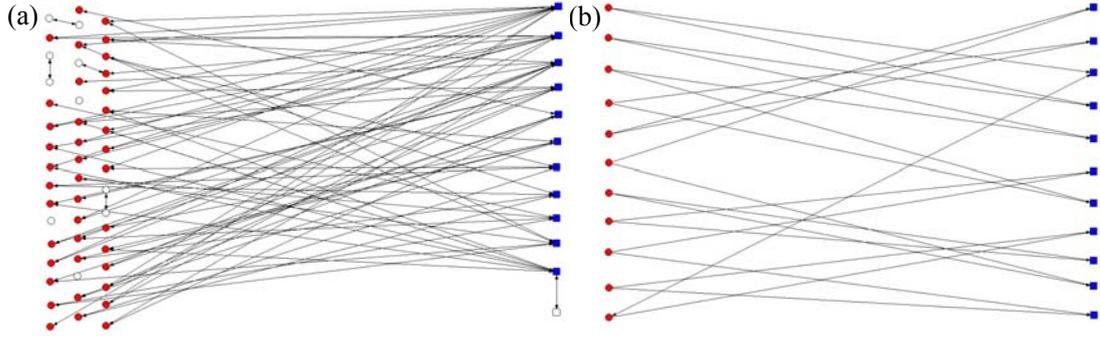

**Fig. 1.** A IERDCN (a) and a special case (b). The red and white circles represent enterprises with and without connecting to institutes, while the blue and white squares represent institutes with and without connecting to enterprises.

Two cases of IERDCNs in China have been investigated in details (see Fig. 1 for their typologies and Table 1 for their characteristics). Both of enterprises and institutes are located in Beijing Zhongguancun Science Park. There are 50 electronics enterprises compose of enterprise R&D network, and 12 institutes compose of institute R&D network in IERDCN 1 (Fig. 1(a)). Similarly, there are 11 electronics enterprises and 11 institutes in IERDCN 2 (Fig. 1(b)). Some large conglomerates such as Lenovo, Digital China, and BOE are very famous for independently R&D, and ones of them like Founder, Tsinghua Tongfang, Datang and Potevio, etc. have formed strategic R&D partnerships with other enterprises that have the long-term business connection, so there are enterprises without collaboration with any institute in institute R&D network. It should be stressed that the links come from patent application, which is an important R&D collaboration results. If a couple of partners have applied for patent protection for a new technique, R&D collaboration between them has become inevitable. All the patent information can be searched from the website of Chinese State Intellectual Property Office.

**Table 1** Characteristics of IERDCN 1 and 2

| IERDCN | 1 | 2 |
| --- | --- | --- |
| Density | 0.0397 | 0.0952 |
| Ave degree | 2.4194 | 2.0000 |
| Ave closeness | 0.0012 | 0.0083 |
| Ave distance | 3.8631 | 5.7619 |
| Ave betweenness | 63.7742 | 50.0000 |
| Ave clustering coefficient | 0.0000 | 0.0000 |

IERDCN 2 extracted from IERDCN 1 is a special case which is not very easy to find an identical network. Actually, it is a classic IERDCN, where has no inter-network R&D collaboration but only two partners coming from the opposite network. This occurs when all enterprises in the same industry have equal scale and market position, as well as there are always unsatisfied demands for future R&D. Furthermore, enterprises prefer to collaborate with two institutes for ensuring higher probability of success and lower cost. Every research institute, in the meantime, has the same powerful technology strength and is ready for collaborative R&D. It is important that equal amounts of collaborations ensure whole network working successfully.

**Table 2** Average intra- and inter- degree of IERDCN 1 and 2

| IERDCN | | 1 | 2 |
| --- | --- | --- | --- |
| Enterprise R&D network | $\bar{d}_{intra-}$ | 0.1600 | 0.0000 |
| | $\bar{d}_{inter-}$ | 1.4000 | 2.0000 |
| Institute R&D network | $\bar{d}_{intra-}$ | 0.1667 | 0.0000 |
| | $\bar{d}_{inter-}$ | 5.8333 | 2.0000 |

Table 2 shows the average intra-network and inter-network degree of IERDCN 1 and 2. In addition to the illustration of Fig. 1, we can see there are rarely intra-network collaborations. On the contrary, more inter-network collaborations exist in IERDCN 1, where the number of collaborations between institutes and enterprises is relatively large. And it is obvious in IERDCN 2, where have equal inter-network collaborations both in the enterprise R&D network and institute R&D network.

We have to explain these structural properties based on the management attributes of enterprises and institutes. Firstly, collaboration and competition coexist in the respective network. Besides similar or identical products and services, competition comes from information, skill and knowledge barrier. Collaboration relies on complementarities in creating common benefits to alliance partners, and it can help to diffuse advanced knowledge and techniques. However, due to the lack of profit-driven and information communication, there are only a few collaborations occur when enterprises or institutes have the same stockholders. Furthermore, we can find that enterprises enjoy collaborating with more than one institutes because the more collaboration the better the chances of generating new techniques and knowledge. Moreover, some institutes have large quantities of partners because of higher R&D capability and reputation. Even so, too many partners of an institute may bring overburden and even lead to failure. In order to avoid excessive competition, much regular collaboration has emerged after several years' competition and collaboration.

We must emphasize that despite there exists a link, does not mean that knowledge and techniques can be able to transmit for certain. The reason may be related to the organizational otherness and differences in absorbency. In practice, especially, the collaborations between enterprises and institutes or between themselves may be failed, or they may accomplish the R&D task of partners selected rather than complete all tasks. Any enterprise and institute may not be able to gain the knowledge or technique. We will introduce connection probability to the modeling process for studying this issue. In doing so, we implicitly assume that each collaboration has the risk of failing to generate knowledge and technique, which are a departure from prior studies that 100 percent successful assumption. Nevertheless, watching the R&D collaborations for eight years allowed us to confirm our assumption and study based on they are reasonable in practice. Additionally, the actual connection probability of intra-networks and inter-networks can be determined by investigation and research. There may be some structural properties reflect the economic, management or social attributes of networks, which should be taken into mathematical models constructing process. And the second arithmetic in the next section will present the corresponding work.

## 3. A mathematical framework based on generating functions

### 3.1. The generating functions of classic IERDCNs

Here we give a brief description of generating functions, which have been used in many network connectivity studies [8-10, 16, 18-21], etc., all found that the functions are quite accurate when the structure of networks is approximately tree-like. Now consider classic IERDCNs formed by two interacting networks, 1 and 2, whose characteristic is very rare to observe nodes in the same network with inner links. The network 1 is composed of enterprises with the similar products and services. Moreover, the network 2 is composed of institutes with identical R&D capability. In practice, it represents a circumstance full of intense competitions and no collaborations between enterprises or institutes in their own network. Each individual network $\mu$ could be characterized by a multi-degree distribution, $\{p_{k_1 k_2}^{\mu}\}$, where $p_{k_1 k_2}^{\mu}$ is the probability that a network $\mu$ node which has $k_1$ edges to nodes in network 1, and $k_2$ edges to nodes in network 2. The multi-degree distribution for network $\mu$ could be written in the form of a generating function:

$$G_\mu(x_1, x_2) = \sum_{k_1=0, k_2=0}^{\infty} p_{k_1 k_2}^{\mu} x_1^{k_1} x_2^{k_2}. \tag{1}$$

We also assume the distribution $p_{k_1 k_2}^{\mu}$ is correctly normalized, so that

$$G_\mu(1, 1) = 1. \tag{2}$$

Firstly, consider selecting uniformly at random a $v$-$\mu$ edge which is used to connect a couple nodes in network $\mu$ and network $v$. Relative to a single network, the remaining local connectivity to nodes in other networks is also accounted by excess degree [19]. We use $p_{k_1 k_2}^{\mu v}$ to denote the probability that randomly chosen $v$-$\mu$ edge to a node with excess $v$ degree which has total $v$-degree of $k_v$+1. Then the generating function for the distribution, $\{p_{k_1 k_2}^{\mu v}\}$ is,

$$G_{\mu\nu}(x) = \sum_{k_1,\ldots,k_l=0}^{\infty} p_{k_1\ldots k_l}^{\mu\nu} x_1^{k_1} \ldots x_l^{k_l}$$

$$= \sum_{k_1=0, k_2=0}^{\infty} \frac{p_{k_1(k_\nu+1)k_2}^{\mu}(k_\nu+1)}{\sum_{j_1=0, j_2=0}^{\infty} p_{j_1(j_\nu+1)j_2}^{\mu}(j_\nu+1)} x_1^{k_1} x_2^{k_2}. \tag{3}$$

According to the structure properties of network 1 and 2 in IERDCN 2, we have $p_{k_1 k_2}^{11} = p_{k_1 k_2}^{22} = 0$, which means $G_{11}(x_1, x_2) = G_{22}(x_1, x_2) = 0$. Thus, we only consider the excess degree of a node in network 1 and 2, actually the generating functions could be written as

$$G_{12}(x_1, x_2) = \sum_{k_1=0, k_2=0}^{\infty} \frac{p_{k_1(k_2+1)}^{1}(k_2+1)}{\sum_{k_1=0, k_2=0}^{\infty} p_{k_1 k_2}^{1} k_2} x_1^{k_1} x_2^{k_2}. \tag{4}$$

$$G_{21}(x_1, x_2) = \sum_{k_1=0, k_2=0}^{\infty} \frac{p_{(k_1+1)k_2}^{2}(k_1+1)}{\sum_{k_1=0, k_2=0}^{\infty} p_{k_1 k_2}^{2} k_1} x_1^{k_1} x_2^{k_2}. \tag{5}$$

Secondly, consider the component sizes of IERDCN 2. All the component sizes are finite in the beginning, after the emergence of a giant connected component, which become larger and larger. While the component sizes are too large to tolerate to ignore the closed loop of edges, the generating functions could be suited to calculate the average component size [8, 16, 18]. They also could be suited to calculate the probability that a randomly chosen node belongs to the giant component in a supercritical regime [8, 16, 18]. We have to scrutinize components of a randomly chosen undirected edge. Let $H_{\nu\mu}(x_1, x_2)$ be the generating function for the distribution of the sizes of components reached by following randomly chosen edges connecting nodes in network $\nu$ with nodes in the network $\mu$.

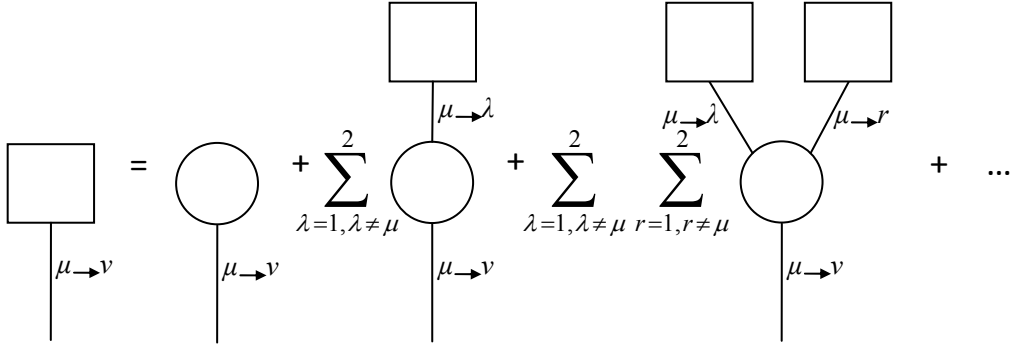

**Fig. 2.** A diagrammatical representation of the topological constraints placed on the generating function $H_{\mu\nu}(x_1, x_2)$ for the distribution of sizes of components reachable by follow a randomly chosen $\nu$-$\mu$ edge. Where $G_{11}(x_1, x_2)=0$ and $G_{22}(x_1, x_2)=0$, so the functions may be expressed as this.

Fig. 2 shows all the types of connectivity possible for the $\mu$ node as the receiver of a randomly chosen edge in its component. Because of $G_{11}(x_1, x_2) = G_{22}(x_1, x_2) = 0$, when there is no giant component, we have:

$$\begin{cases} H_{12}(x_1, x_2) = x_1 \cdot G_{12}[H_{11}(x_1, x_2), H_{21}(x_1, x_2)] \\ H_{21}(x_1, x_2) = x_2 \cdot G_{21}[H_{12}(x_1, x_2), H_{22}(x_1, x_2)]. \end{cases} \tag{6}$$

By taking partial derivative of both sides in each sub-equation of Eq. (6) with respect to $x_\mu$, the average component size for any $H_{\mu\nu}(x_1, x_2)$ could be calculated. Then, let $x_1=1$ and $x_2=1$ and put them into the calculating progress. For the sake of briefness, we would like to use $H_\mu^{'\lambda}(1,1)$, $H_{\mu\nu}^{'\lambda}(1,1)$, $G_\mu^{'\lambda}(1,1)$ and $G_{\mu\nu}^{'\lambda}(1,1)$ instead of $\left.\frac{\partial H_\mu(x_1,x_2)}{\partial x_\lambda}\right|_{x_1=1, x_2=2}$, $\left.\frac{\partial H_{\mu\nu}(x_1,x_2)}{\partial x_\lambda}\right|_{x_1=1, x_2=2}$, $\left.\frac{\partial G_\mu(x_1,x_2)}{\partial x_\lambda}\right|_{x_1=1, x_2=2}$ and $\left.\frac{\partial G_{\mu\nu}(x_1,x_2)}{\partial x_\lambda}\right|_{x_1=1, x_2=2}$ in the following parts of this paper when there is no ambiguity. Thus

$$\begin{cases} H_{12}^{'1}(1,1) = 1 + G_{12}^{'1}(1,1) \cdot H_{11}^{'1}(1,1) + G_{12}^{'2}(1,1) \cdot H_{21}^{'1}(1,1) \\ H_{21}^{'1}(1,1) = G_{21}^{'1}(1,1) \cdot H_{12}^{'1}(1,1) + G_{21}^{'2}(1,1) \cdot H_{22}^{'1}(1,1) \\ H_{12}^{'2}(1,1) = G_{12}^{'1}(1,1) \cdot H_{11}^{'2}(1,1) + G_{12}^{'2}(1,1) \cdot H_{21}^{'2}(1,1) \\ H_{21}^{'2}(1,1) = 1 + G_{21}^{'2}(1,1) \cdot H_{12}^{'2}(1,1) + G_{21}^{'1}(1,1) \cdot H_{22}^{'2}(1,1). \end{cases} \tag{7}$$

In Eq. (7) all $G_{\mu\nu}^{'\lambda}(1,1)$ could be calculated based on Eqs. (4) and (5), if we have known the multi-degree distribution of each network. For instance, $G_{12}^{'1}(1,1) = \frac{\overline{k_{12}} \cdot \overline{k_{11}}}{\overline{k_{12}}}$, where $k_{11}$ and $k_{12}$ have the same meanings with $k_1$ and $k_2$ in Eq. (1) and Eq. (3). Meanwhile, all $H_{\mu\nu}^{'\lambda}(1,1)$ could be solved in the Eq. (7), which has four sub-equations and four unknowns.

The generating function for the probability distribution of component size of a randomly selected $\mu$ node when there is no giant component can be written as $H_\mu(x_1, x_2) = x_\mu [H_{1\mu}(x_1, x_2), H_{\mu 2}(x_1, x_2)]$, we have

$$\begin{cases} H_1^{'2}(1,1) = G_1^{'1}(1,1) \cdot H_{11}^{'2}(1,1) + G_1^{'2}(1,1) \cdot H_{21}^{'2}(1,1) \\ H_2^{'1}(1,1) = G_2^{'1}(1,1) \cdot H_{12}^{'1}(1,1) + G_2^{'2}(1,1) \cdot H_{22}^{'1}(1,1). \end{cases} \quad (8)$$

While the $H_\mu^{'\lambda}(1,1)$ could be calculated by putting the results of Eq. (7) into Eq. (8). Now the average component size will be solved.

Thirdly, we would like to discuss the percolation threshold. Because of no inner links in both networks, $H_{11}^{'1}(1,1)$, $H_{22}^{'1}(1,1)$, $H_{11}^{'2}(1,1)$ and $H_{22}^{'2}(1,1)$ should be canceled. For that reason, we can conclude that $G_{12}^{'1}(1,1)$ and $G_{21}^{'1}(1,1)$ equals 0 (because of $k_{11}=k_{22}=0$). Thus, $H_{12}^{'2}(1,1)$ and $H_{21}^{'2}(1,1)$ equals 0. Let's plug $G_{11}^{'1}(1,1) = G_{11}^{'2}(1,1) = G_{22}^{'1}(1,1) = G_{22}^{'2}(1,1) = 0$ into Eq. (7), and it could be written as

$$\begin{cases} H_{12}^{'1}(1,1) = 1 + G_{12}^{'2}(1,1) \cdot H_{21}^{'1}(1,1) \\ H_{21}^{'1}(1,1) = G_{21}^{'1}(1,1) \cdot H_{12}^{'1}(1,1). \end{cases} \quad (9)$$

Solving Eq. (9) and we get

$$\begin{cases} H_{12}^{'1}(1,1) = \dfrac{1}{1 - G_{21}^{'1}(1,1) \cdot G_{12}^{'2}(1,1)} \\ H_{21}^{'1}(1,1) = \dfrac{G_{21}^{'1}(1,1)}{1 - G_{21}^{'1}(1,1) \cdot G_{12}^{'2}(1,1)}. \end{cases} \quad (10)$$

We believe that the necessary and sufficient condition for the absence of the giant component of IERDCN 2 is

$$G_{12}^{'2}(1,1) \cdot G_{21}^{'1}(1,1) < 1. \quad (11)$$

Finally, let us calculate the probability that a randomly chosen node belongs to the giant component. Here $u_{\mu\nu}$ represents the probability that a randomly chosen edge pointing at a $\mu$ node leaving from a network $\nu$ node is not part of the giant component. We get

$$\begin{cases} u_{12} = G_{12}(u_{11}, u_{21}) \\ u_{21} = G_{21}(u_{12}, u_{22}). \end{cases} \quad (12)$$

Taking $G_{12}$ and $G_{21}$ as known, we can solve $u_{12}$ and $u_{21}$, then we can calculate the probability that a randomly chosen $\mu$ node belongs to the giant component $S_\mu$ as follows:

$$\begin{cases} S_1 = 1 - G_1(u_{11}, u_{21}) \\ S_2 = 1 - G_2(u_{12}, u_{22}). \end{cases} \quad (13)$$

Above all, we discuss generating functions of the classic IERDCNs. In the following, we will show that our arithmetic based on connection probability.

### 3.2. The generating functions of IERDCNs with connection probability

Now we are in the position to introduce the case with connection probability $\omega$, where collaborations achieve partial success in IERDCNs. Before discussing, we use $\omega_\mu$, $\omega_\nu$ and $\omega_{\mu\nu}$ to denote separately the threshold value of connecting a couple nodes in intra-networks and inter-network. In practice, the collaboration will fail to maintain if does not generate knowledge or technique, which cannot be able to flow in networks. Nevertheless, many researchers believe that the collaboration is succeeded (links are existed) when enterprises try to contact institutes even if have signed collaborative agreements. For better to describe this complex and dynamic phenomenon, we introduce connection probability, which represents the probability of succeed in collaborating. According to the different situation, the connection probability of enterprises and institutes may be different. But now they have equal collaboration choice because rarely inner links existed, so we let $\omega_\mu$ and $\omega_\nu$ equal $\omega_1$, $\omega_{\mu\nu}$ equal $\omega_2$. Based on generating functions mentioned above, we also use $G_\mu$ to calculate the multi-degree distribution.

Firstly, connection probability was used to endow different weight for every $\nu$-$\mu$ edge, means it may break some couple nodes selected randomly from network $\mu$ and network $v$. Furthermore, the optimal target range of the connection probability is from 0 to $\omega_i$, if beyond, the connection will break. So by using probability that randomly chosen $v$-$\mu$ edge to a node with excess $v$ degree, the generating functions of IERDCNs with connection probability based Eq. (3) could be written as

$$\begin{cases} G_{11}(\omega_1 \cdot x_1, \omega_2 \cdot x_2) = \sum_{k_1=0,k_2=0}^{\infty} \dfrac{p^1_{(k_1+1)k_2}(k_1+1)}{\sum_{k_1=0,k_2=0}^{\infty} p^1_{k_1 k_2} k_1} \cdot (\omega_1 \cdot x_1)^{k_1} \cdot (\omega_2 \cdot x_2)^{k_2} \\[2mm] G_{12}(\omega_1 \cdot x_1, \omega_2 \cdot x_2) = \sum_{k_1=0,k_2=0}^{\infty} \dfrac{p^1_{k_1(k_2+1)}(k_2+1)}{\sum_{k_1=0,k_2=0}^{\infty} p^1_{k_1 k_2} k_2} \cdot (\omega_1 \cdot x_1)^{k_1} \cdot (\omega_2 \cdot x_2)^{k_2} \\[2mm] G_{21}(\omega_1 \cdot x_1, \omega_2 \cdot x_2) = \sum_{k_1=0,k_2=0}^{\infty} \dfrac{p^2_{(k_1+1)k_2}(k_1+1)}{\sum_{k_1=0,k_2=0}^{\infty} p^2_{k_1 k_2} k_1} \cdot (\omega_1 \cdot x_1)^{k_1} \cdot (\omega_2 \cdot x_2)^{k_2} \\[2mm] G_{22}(\omega_1 \cdot x_1, \omega_2 \cdot x_2) = \sum_{k_1=0,k_2=0}^{\infty} \dfrac{p^2_{k_1(k_2+1)}(k_2+1)}{\sum_{k_1=0,k_2=0}^{\infty} p^2_{k_1 k_2} k_2} \cdot (\omega_1 \cdot x_1)^{k_1} \cdot (\omega_2 \cdot x_2)^{k_2}. \end{cases} \quad (14)$$

The multi-degree distribution for each network may be written in the form of a generating function:

$$\begin{cases} G_1(\omega_1 \cdot x_1, \omega_2 \cdot x_2) = \sum_{k_1=0,k_2=0}^{\infty} p^1_{k_1 k_2} \cdot (\omega_1 \cdot x_1)^{k_1} \cdot (\omega_2 \cdot x_2)^{k_2} \\ G_2(\omega_1 \cdot x_1, \omega_2 \cdot x_2) = \sum_{k_1=0,k_2=0}^{\infty} p^2_{k_1 k_2} \cdot (\omega_1 \cdot x_1)^{k_1} \cdot (\omega_2 \cdot x_2)^{k_2}. \end{cases} \quad (15)$$

Secondly, we discuss the component sizes. We also have to scrutinize components of a randomly chosen undirected edge with connection probability. Let $H_{v\mu}(x_1, x_2)$ be the new generating function for the distribution of the sizes of components reached by following randomly chosen $v$-$\mu$ edges in the optimal target interval. A key point that needs to be emphasized is that the links between the $v$-$\mu$ nodes are one hundred percent exist.

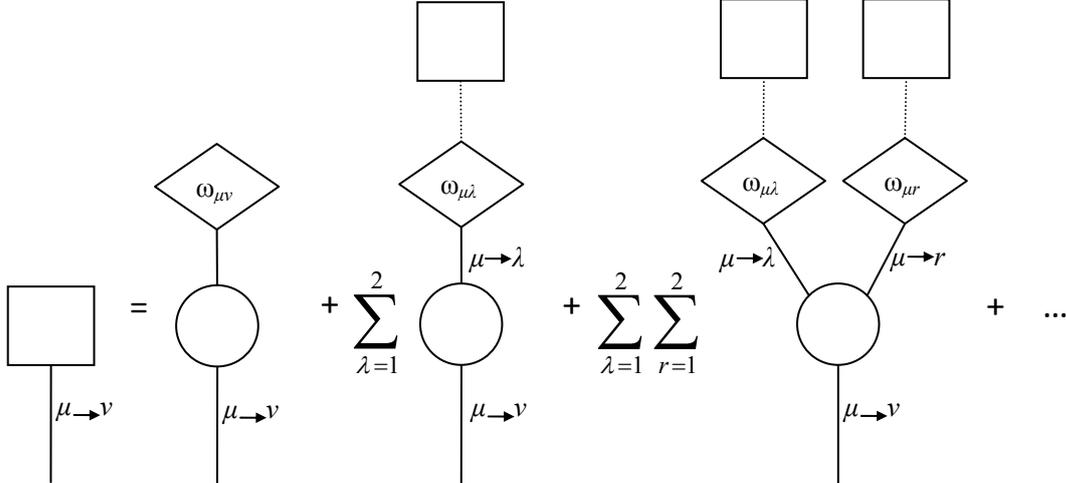

**Fig. 3.** A diagrammatical representation of the topological constraints placed on the generating function $H_{\mu\nu}(\omega_1 \cdot x_1, \omega_2 \cdot x_2)$ for the distribution of sizes of components reachable by follow a randomly chosen $v$-$\mu$ edge, where connection probability decides all links in IERDCNs.

Fig. 3 shows all the types of connectivity possible for the $\mu$ node as the receiver of a randomly chosen edge with some connection probability in its component. Meanwhile, let $\omega_{\lambda\mu} \cdot H_{\lambda\mu}(x_1,x_2)^{k_1} + (1-\omega_{\lambda\mu}) \cdot H_{\lambda\mu}(x_1,x_2)^0$ denotes the generating function for the distribution of the sizes of components with total probability, we explicitly exclude from $H_{\lambda\mu}(x_1, x_2)$ the giant component and get

$$H_{\mu\nu}(x_1, x_2) = x_\mu \cdot \sum_{k_1=0,k_2=0}^{\infty} p^{\mu\nu}_{k_1 k_2} \cdot [\omega_{1\mu} \cdot H_{1\mu}(x_1,x_2)^{k_1} + 1 - \omega_{1\mu}, \omega_{\lambda\mu} \cdot H_{1\mu}(x_1,x_2)^{k_2} + 1 - \omega_{\lambda\mu}]. \quad (16)$$

where if $\mu=\lambda$, $\omega_{\mu\nu}=\omega_1$, else $\omega_{\mu\nu}=\omega_2$. We could change the form of Eq. (16) from Eq. (14) and Eq. (15). Thus

$$H_{\mu\nu}(x_1, x_2) = x_\mu \cdot G_{\mu\nu}[\omega_{1\mu} \cdot H_{1\mu}(x_1,x_2) + 1 - \omega_{1\mu}, \omega_{\lambda\mu} \cdot H_{\lambda\mu}(x_1,x_2) + 1 - \omega_{\lambda\mu}]. \quad (17)$$

Considering starting from a randomly chosen $\mu$ node, instead of $v$-$\mu$ edge, we can get a topology describes the end of each edge incident to the $\mu$ node, such as one from Fig. 3. The generating function represents probability distribution of component sizes could be written as
$$H_\mu(x_1, x_2) = x_\mu \cdot G_\mu[\omega_{1\mu} \cdot H_{1\mu}(x_1, x_2) + 1 - \omega_{1\mu}, \omega_{\lambda\mu} \cdot H_{\lambda\mu}(x_1, x_2) + 1 - \omega_{\lambda\mu}]. \tag{18}$$

In IERDCNs, the Eq. (17) becomes
$$\begin{cases} H_{11}(x_1, x_2) = x_1 \cdot G_{11}[\omega_1 \cdot H_{11}(x_1, x_2) + 1 - \omega_1, \omega_2 \cdot H_{21}(x_1, x_2) + 1 - \omega_2] \\ H_{12}(x_1, x_2) = x_1 \cdot G_{12}[\omega_1 \cdot H_{11}(x_1, x_2) + 1 - \omega, \omega_2 \cdot H_{21}(x_1, x_2) + 1 - \omega_2] \\ H_{21}(x_1, x_2) = x_2 \cdot G_{21}[\omega_2 \cdot H_{12}(x_1, x_2) + 1 - \omega_2, \omega_1 \cdot H_{22}(x_1, x_2) + 1 - \omega] \\ H_{22}(x_1, x_2) = x_2 \cdot G_{22}[\omega_2 \cdot H_{12}(x_1, x_2) + 1 - \omega_2, \omega_1 \cdot H_{22}(x_1, x_2) + 1 - \omega]. \end{cases} \tag{19}$$

And the Eq. (18) becomes
$$\begin{cases} H_1(1,1) = x_1 \cdot G_1[\omega_1 \cdot H_{11}(1,1) + 1 - \omega_1, \omega_2 \cdot H_{21}(1,1) + 1 - \omega_2] \\ H_2(1,1) = x_2 \cdot G_2[\omega_2 \cdot H_{12}(1,1) + 1 - \omega_2, \omega_1 \cdot H_{22}(1,1) + 1 - \omega_1]. \end{cases} \tag{20}$$

Take partial derivative of both sides in each sub-equation of Eqs. (14)-(20) with respect to $x_1$ and $x_2$. Then, make $x_1=1$ and $x_2=1$ and put them into the calculating progress. We have
$$\begin{cases} H_{11}^{'1}(1,1) = 1 + \omega_1 \cdot G_{11}^{'1}(1,1) \cdot H_{11}^{'1}(1,1) + \omega_2 \cdot G_{11}^{'2}(1,1) \cdot H_{21}^{'1}(1,1) \\ H_{12}^{'1}(1,1) = 1 + \omega_1 \cdot G_{12}^{'1}(1,1) \cdot H_{11}^{'1}(1,1) + \omega_2 \cdot G_{12}^{'2}(1,1) \cdot H_{21}^{'1}(1,1) \\ H_{21}^{'1}(1,1) = \omega_2 \cdot G_{21}^{'1}(1,1) \cdot H_{12}^{'1}(1,1) + \omega_1 \cdot G_{21}^{'2}(1,1) \cdot H_{22}^{'1}(1,1) \\ H_{22}^{'1}(1,1) = \omega_2 \cdot G_{22}^{'1}(1,1) \cdot H_{12}^{'1}(1,1) + \omega_1 \cdot G_{22}^{'2}(1,1) \cdot H_{22}^{'1}(1,1) \\ H_{11}^{'2}(1,1) = \omega_1 \cdot G_{11}^{'2}(1,1) \cdot H_{11}^{'2}(1,1) + \omega_2 \cdot G_{11}^{'1}(1,1) \cdot H_{21}^{'2}(1,1) \\ H_{12}^{'2}(1,1) = \omega_1 \cdot G_{12}^{'2}(1,1) \cdot H_{11}^{'2}(1,1) + \omega_2 \cdot G_{12}^{'1}(1,1) \cdot H_{21}^{'2}(1,1) \\ H_{21}^{'2}(1,1) = 1 + \omega_2 \cdot G_{21}^{'2}(1,1) \cdot H_{12}^{'2}(1,1) + \omega_1 \cdot G_{21}^{'1}(1,1) \cdot H_{22}^{'2}(1,1) \\ H_{12}^{'1}(1,1) = 1 + \omega_2 \cdot G_{22}^{'2}(1,1) \cdot H_{12}^{'2}(1,1) + \omega_1 \cdot G_{22}^{'1}(1,1) \cdot H_{22}^{'2}(1,1). \end{cases} \tag{21}$$

and
$$\begin{cases} H_1^{'1}(1,1) = 1 + G_1^{'1}(1,1) \cdot \omega_1 \cdot H_{11}^{'1}(1,1) + G_1^{'2}(1,1) \cdot \omega_2 \cdot H_{21}^{'1}(1,1) \\ H_1^{'2}(1,1) = G_1^{'1}(1,1) \cdot \omega_1 \cdot H_{11}^{'2}(1,1) + G_1^{'2}(1,1) \cdot \omega_2 \cdot H_{21}^{'2}(1,1) \\ H_2^{'1}(1,1) = G_2^{'1}(1,1) \cdot \omega_2 \cdot H_{12}^{'1}(1,1) + G_2^{'2}(1,1) \cdot \omega_1 \cdot H_{22}^{'1}(1,1) \\ H_2^{'2}(1,1) = 1 + G_2^{'1}(1,1) \cdot \omega_2 \cdot H_{12}^{'2}(1,1) + G_2^{'2}(1,1) \cdot \omega_1 \cdot H_{22}^{'2}(1,1). \end{cases} \tag{22}$$

While the $H_\mu^{'\lambda}(1,1)$ could be calculated by putting the results of Eq. (21) into Eq. (22). Now the average component size will be solved.

Thirdly, we would like to discuss the percolation threshold. By solving Eq. (21) we could get

$$\begin{cases}
H_{11}^{'1}(1,1) = \dfrac{1 - \omega_1 \cdot G_{22}^{'2}(1,1) + \omega_2^{\,2} \cdot (G_{11}^{'2}(1,1) - G_{12}^{'2}(1,1))}{\theta} \\
\qquad \cdot [G_{21}^{'1}(1,1) \cdot (1 - \omega_1 \cdot G_{22}^{'2}(1,1)) + \omega_1 \cdot G_{21}^{'1}(1,1) \cdot G_{22}^{'1}(1,1)] \\[4pt]
H_{12}^{'1}(1,1) = \dfrac{(1 - \omega_1 \cdot G_{22}^{'2}(1,1)) \cdot (1 - \omega_1 \cdot G_{11}^{'1}(1,1) + \omega_1 \cdot G_{12}^{'1}(1,1))}{\theta} \\[4pt]
H_{21}^{'1}(1,1) = \dfrac{(1 - \omega_1 \cdot G_{11}^{'1}(1,1) + \omega_1 \cdot G_{12}^{'1}(1,1)) \cdot \omega_2 \cdot [G_{21}^{'1}(1,1) \cdot (1 - \omega_1 \cdot G_{22}^{'2}(1,1)) + \omega_1 \cdot G_{21}^{'1}(1,1) \cdot G_{22}^{'1}(1,1)]}{\theta} \\[4pt]
H_{22}^{'1}(1,1) = \dfrac{G_{22}^{'1}(1,1) \cdot \omega_2 \cdot (1 - \omega_1 \cdot G_{11}^{'1}(1,1) + \omega_1 \cdot G_{12}^{'1}(1,1))}{\theta} \\[4pt]
H_{11}^{'2}(1,1) = \dfrac{G_{11}^{'1}(1,1) \cdot \omega_2 \cdot (1 - \omega_1 \cdot G_{22}^{'2}(1,1) + \omega_1 \cdot G_{21}^{'1}(1,1))}{\sigma} \\[4pt]
H_{12}^{'2}(1,1) = \dfrac{\omega_2 \cdot (1 - \omega_1 \cdot G_{22}^{'1}(1,1) + \omega_1 \cdot G_{21}^{'1}(1,1)) \cdot [G_{12}^{'1}(1,1) \cdot (1 - \omega_1 \cdot G_{11}^{'2}(1,1)) + \omega_1 \cdot G_{11}^{'1}(1,1) \cdot G_{12}^{'2}(1,1)]}{\sigma} \\[4pt]
H_{21}^{'2}(1,1) = \dfrac{(1 - \omega_1 \cdot G_{11}^{'2}(1,1)) \cdot (1 - \omega_1 \cdot G_{22}^{'1}(1,1) + \omega_1 \cdot G_{21}^{'1}(1,1))}{\sigma} \\[4pt]
H_{22}^{'2}(1,1) = \dfrac{1 - \omega_1 \cdot G_{11}^{'2}(1,1) + \omega_2^{\,2} \cdot (G_{21}^{'2}(1,1) - G_{22}^{'2}(1,1))}{\sigma} \\
\qquad \cdot [G_{12}^{'1}(1,1) \cdot (\omega_1 \cdot G_{11}^{'2}(1,1) - 1) - \omega_1 \cdot G_{11}^{'1}(1,1) \cdot G_{12}^{'2}(1,1)].
\end{cases}$$ 
(23)

where

$$\begin{aligned}
\theta =\ & (1 - \omega_1 \cdot G_{11}^{'1}(1,1)) \cdot (1 - \omega_1 \cdot G_{22}^{'2}(1,1)) \cdot (1 - \omega_2 \cdot G_{12}^{'1}(1,1) \cdot \omega_2 \cdot G_{21}^{'1}(1,1)) \\
& - \omega_1 \cdot G_{11}^{'2}(1,1) \cdot \omega_2 \cdot G_{12}^{'1}(1,1) \cdot \omega_2 \cdot G_{21}^{'1}(1,1) \cdot (1 - \omega_1 \cdot G_{22}^{'2}(1,1)) \\
& - \omega_2 \cdot G_{12}^{'2}(1,1) \cdot \omega_2 \cdot G_{21}^{'1}(1,1) \cdot \omega_1 \cdot G_{22}^{'1}(1,1) \cdot (1 - \omega_1 \cdot G_{11}^{'1}(1,1)) \\
& - \omega_1 \cdot G_{11}^{'2}(1,1) \cdot \omega_2 \cdot G_{12}^{'1}(1,1) \cdot \omega_2 \cdot G_{21}^{'2}(1,1) \cdot \omega_1 \cdot G_{22}^{'1}(1,1).
\end{aligned}$$ 
(24)

$$\begin{aligned}
\sigma =\ & (1 - \omega_1 \cdot G_{11}^{'2}(1,1)) \cdot (1 - \omega_1 \cdot G_{22}^{'1}(1,1)) \cdot (1 - \omega_2 \cdot G_{12}^{'1}(1,1) \cdot \omega_2 \cdot G_{21}^{'2}(1,1)) \\
& - \omega_1 \cdot G_{11}^{'1}(1,1) \cdot \omega_2 \cdot G_{12}^{'1}(1,1) \cdot \omega_2 \cdot G_{21}^{'2}(1,1) \cdot (1 - \omega_1 \cdot G_{22}^{'2}(1,1)) \\
& - \omega_2 \cdot G_{12}^{'1}(1,1) \cdot \omega_2 \cdot G_{21}^{'1}(1,1) \cdot \omega_1 \cdot G_{22}^{'2}(1,1) \cdot (1 - \omega_1 \cdot G_{11}^{'2}(1,1)) \\
& - \omega_1 \cdot G_{11}^{'1}(1,1) \cdot \omega_2 \cdot G_{12}^{'1}(1,1) \cdot \omega_2 \cdot G_{21}^{'1}(1,1) \cdot \omega_1 \cdot G_{22}^{'2}(1,1).
\end{aligned}$$ 
(25)

Eq. (23)-(25) show that the necessary and sufficient condition for the absence of the system-wide giant component is

$$\begin{cases} \theta > 0 \wedge \omega_1 \cdot G_{11}^{'1}(1,1) < 1 \wedge \omega_1 \cdot G_{22}^{'2}(1,1) < 1 \wedge \omega_2^{\,2} \cdot G_{12}^{'2}(1,1) \cdot G_{21}^{'1}(1,1) < 1 \\ \sigma > 0 \wedge \omega_1 \cdot G_{11}^{'2}(1,1) < 1 \wedge \omega_1 \cdot G_{22}^{'1}(1,1) < 1 \wedge \omega_2^{\,2} \cdot G_{12}^{'1}(1,1) \cdot G_{21}^{'2}(1,1) < 1. \end{cases}$$ 
(26)

To prove the sufficiency, we apply these conditions to Eq. (23), and find that according Eq. (14) and Eq. (21), $H_{11}^{'1}(1,1) = \dfrac{1 + \omega_2 \cdot G_{11}^{'2}(1,1) \cdot H_{21}^{'1}(1,1)}{1 - \omega_1 \cdot G_{11}^{'1}(1,1)}$, $G_{\mu\nu}^{'}(1,1) \equiv \bar{k}_{\mu\nu} \geq 0$ and $G_{11}^{'1}(1,1)$ almost equal to 0 (because of rarely inner links), so $H_{11}^{'1}(1,1)$ converge to a value equal to or higher than 1. Furthermore, $H_{12}^{'1}(1,1)$ converge to a values equal to or higher than 1, $H_{21}^{'1}(1,1)$ and $H_{22}^{'1}(1,1)$ converge to values equal to or higher than 0. In the same way, $H_{11}^{'2}(1,1)$ and $H_{12}^{'2}(1,1)$ converge to values equal to or higher than 0. Both $H_{21}^{'2}(1,1)$ and $H_{22}^{'2}(1,1)$ converge to value equal to or higher than 1. For all parameters, when Eq. (26) is satisfied, the sufficiency is proved. And when $\omega_1 \cdot G_{11}^{'1}(1,1) \geq 1 \vee \omega_1 \cdot G_{22}^{'2}(1,1) \geq 1 \vee \omega_2 \cdot G_{11}^{'2}(1,1) \geq 1 \vee \omega_2 \cdot G_{22}^{'1}(1,1) \geq 1$, a part of the whole network has already undergone the phase transition. Otherwise, a giant component appears. Considering the sufficiency of Eq. (25) which has been proven, we have it is the necessary and sufficient condition for the absence of the system-wide giant component.

Finally, the probability that randomly chosen nodes and the fraction of network nodes belong to the giant component will be calculated. Once a giant component appears, we can calculate properties of components not belonging to it. As Eq. (12) showing, $u_{\mu\nu}$ represents the probability that a randomly chosen edge pointing at a $\mu$ node leaving from a network $v$ node is not part of the giant component. As we mentioned that whether each edge exists depends on connection probability, our work must in order to ensure that all randomly chosen $v$-$\mu$ edge does not belong to the giant component. Now let $G_{\mu v}$ remains the same, we divide the parameter of $G_{\mu v}$ into two parts. As the probability, both of them are used for calculating the distribution of outgoing edges excluding the giant component. While the connection probability equal to $1-\omega_{\mu v}$, there is neither outgoing edge nor giant component appears in all networks. On the contrary, while the connection probability equal to $\omega_{\mu v}$, which meaning outgoing edge exists and according to Eq. (15), the probability that outgoing edges not belonging to the giant component is $\omega_{\mu v} \cdot u_{\mu v}$. Thus, the probability that a randomly chosen node does not belong to the giant component is $1-\omega_{\mu v}+\omega_{\mu v}\cdot u_{\mu v}$. We get

$$\begin{cases} u_{11} = G_{11}(1-\omega_1+\omega_1 \cdot u_{11}, 1-\omega_2+\omega_2 \cdot u_{21}) \\ u_{12} = G_{12}(1-\omega_1+\omega_1 \cdot u_{11}, 1-\omega_2+\omega_2 \cdot u_{21}) \\ u_{21} = G_{21}(1-\omega_2+\omega_2 \cdot u_{12}, 1-\omega_1+\omega_1 \cdot u_{22}) \\ u_{22} = G_{22}(1-\omega_2+\omega_2 \cdot u_{12}, 1-\omega_1+\omega_1 \cdot u_{22}). \end{cases} \quad (27)$$

and

$$\begin{cases} u_1 = G_1(1-\omega_1+\omega_1 \cdot u_{11}, 1-\omega_2+\omega_2 \cdot u_{21}) \\ u_2 = G_2(1-\omega_2+\omega_2 \cdot u_{12}, 1-\omega_1+\omega_1 \cdot u_{22}). \end{cases} \quad (28)$$

Let $S_\mu$ be the fraction of $\mu$-nodes belonging to the giant component, it may written as

$$\begin{cases} S_1 = 1-u_1 \\ S_2 = 1-u_2. \end{cases} \quad (29)$$

Our discussion up to the case is general and applicable to randomly connecting networks.

## 4. Applications

### 4.1. Comparison with actual networks

In this section, we apply our mathematical framework to the two actual IERDCNs with 100% connection probability mentioned in Section 2. As Table 3 shows, in the first one, there are some discrepancies, but the discrepancies of the theory calculation is proved acceptable. And in the second one, analytical results completely equal to empirical data. We will discuss in more detail, and see how precise our arithmetic is. For IERDCN 1, we get $G_{11}^{'1}(1,1)=0$, $G_{11}^{'2}(1,1)=0.125$, $G_{12}^{'1}(1,1)=0.0142857143$, $G_{12}^{'2}(1,1)=0.8571428571$, $G_{12}^{'2}(1,1)=0.8571428571$, $G_{21}^{'1}(1,1)=6.2$, $G_{21}^{'2}(1,1)=0.1$, $G_{22}^{'1}(1,1)=3.5$, $G_{22}^{'2}(1,1)=0$. Obviously, given that $G_{12}^{'2}(1,1)\cdot G_{21}^{'1}(1,1)=5.3142857140>1$, a giant connected component exists according to Eq. (26) while $\omega_2=1$. Two values of the analytical results and empirical data are extremely close to each other. For IERDCN 2, we get $G_{12}^{'2}(1,1)=1$, $G_{21}^{'1}(1,1)=1$, and $G_{12}^{'2}(1,1)\cdot G_{21}^{'1}(1,1)=1$ which disagree with Eq. (11) and Eq. (26) while $\omega_1=0$ and $\omega_2=1$. Thus, the IERDCN 2 also have a giant connected component.

**Table 3**
Analytical results and empirical data of IERDCN 1 and 2.

|  | IERDCN 1 | | IERDCN 2 | |
| --- | --- | --- | --- | --- |
|  | $S_1$ | $S_2$ | $S_1$ | $S_2$ |
| Analytical results | 0.8174 | 0.9582 | 1.0000 | 1.0000 |
| Empirical data | 0.8200 | 1.0000 | 1.0000 | 1.0000 |

**Table 4**
Analytical results and empirical data of IERDCN 1.

|  | When $\omega_1=0.1$ and $\omega_2=0.1$ | | When $\omega_1=0.1$ and $\omega_2=0.2$ | |
| --- | --- | --- | --- | --- |
| Enterprise R&D network | $H_1^{'1}(1,1)$ | $H_1^{'2}(1,1)$ | $H_1^{'1}(1,1)$ | $H_1^{'2}(1,1)$ |
| Analytical results | 1.0396 | 0.1172 | 1.1906 | 0.2345 |
| Empirical data (average) | 1.0250 | 0.1200 | 1.1735 | 0.2285 |
| Institute R&D network | $H_2^{'1}(1,1)$ | $H_2^{'2}(1,1)$ | $H_2^{'1}(1,1)$ | $H_2^{'2}(1,1)$ |
| Analytical results | 0.5202 | 1.0014 | 1.2526 | 1.0072 |
| Empirical data (average) | 0.5000 | 1.0000 | 1.2417 | 1.0000 |

However, according to our arithmetic, we give two cases with smaller connection probability, let $\omega_1=0.1$, $\omega_2=0.1$ and $\omega_1=0.1$, $\omega_2=0.2$ for IERDCN 1, we can get

$$\begin{cases} G_{11}^{'1}(1,1) = 0 \\ G_{11}^{'2}(1,1) = 0.125 \\ G_{12}^{'1}(1,1) = 0.0143 \\ G_{12}^{'2}(1,1) = 0.8571 \\ G_{21}^{'1}(1,1) = 6.2 \\ G_{21}^{'2}(1,1) = 0.1 \\ G_{22}^{'1}(1,1) = 3.5 \\ G_{22}^{'2}(1,1) = 0 \end{cases} \text{and} \begin{cases} H_{11}^{'1}(1,1) = 1.0083 \\ H_{11}^{'2}(1,1) = 0 \\ H_{12}^{'1}(1,1) = 1.0580 \\ H_{12}^{'2}(1,1) = 0.0028 \\ H_{21}^{'1}(1,1) = 0.6597 \\ H_{21}^{'2}(1,1) = 1.9539 \\ H_{22}^{'1}(1,1) = 0.3703 \\ H_{22}^{'2}(1,1) = 1.5385. \end{cases}, \begin{cases} G_{11}^{'1}(1,1) = 0 \\ G_{11}^{'2}(1,1) = 0.125 \\ G_{12}^{'1}(1,1) = 0.0143 \\ G_{12}^{'2}(1,1) = 0.8571 \\ G_{21}^{'1}(1,1) = 6.2 \\ G_{21}^{'2}(1,1) = 0.1 \\ G_{22}^{'1}(1,1) = 3.5 \\ G_{22}^{'2}(1,1) = 0 \end{cases} \text{and} \begin{cases} H_{11}^{'1}(1,1) = 1.0397 \\ H_{11}^{'2}(1,1) = 0 \\ H_{12}^{'1}(1,1) = 1.2738 \\ H_{12}^{'2}(1,1) = 0.0084 \\ H_{21}^{'1}(1,1) = 1.5841 \\ H_{21}^{'2}(1,1) = 1.9541 \\ H_{22}^{'1}(1,1) = 0.8916 \\ H_{22}^{'2}(1,1) = 1.5385. \end{cases}$$

at this point, we have

$$\theta = 1 > 0 \wedge G_{11}^{'1}(1,1) = 0 < 1 \wedge G_{22}^{'2}(1,1) = 0 < 1 \wedge 0.1 \cdot G_{12}^{'2}(1,1) \cdot 0.1 \cdot G_{21}^{'1}(1,1) = 0.0531 < 1$$

and

$$\theta = 1 > 0 \wedge G_{11}^{'1}(1,1) = 0 < 1 \wedge G_{22}^{'2}(1,1) = 0 < 1 \wedge 0.2 \cdot G_{12}^{'2}(1,1) \cdot 0.2 \cdot G_{21}^{'1}(1,1) = 0.2126 < 1$$

Since both of them satisfy Eq. (26), there is no giant component. Now we can calculate the average component size of a randomly selected μ. By taking into account all possible situations, we work out the average true values of component size. Table 4 shows the analytical values and true values of component size. We find all pairs have tolerated discrepancy (approximately 0.67%~1.4%) which is lower than tolerance of 1 node. The discrepancy may come from the inherent inaccuracy of the generating function. But, it is enough precise to estimate whether the networks have undergone the phase transition and calculate important structural property measures.

*4.2. Application in actual networks*

All the work we do possess reclamation of theory meaning, and also has a very strong practice meaning.

First of all, the component of nodes utilizes resources of the networks most efficiently by inside competing and outside collaborating, which provides a path that every node in the component could learn some technique or knowledge from others without working directly together. As Table 3 shows, in IERDCN 2, every enterprise and institute have a path to the giant component. Original technological innovation ideas may come from every enterprise and institute and must be done in partnership. And for links within the R&D network of enterprises or institutes, it can be found that even if we remove all of them, the network still has a giant component. It means that necessary knowledge of technology R&D can free flow in the whole network. And there is a high success rate of patent output.

Secondly, we introduce connection probability for extending the application of generating function. Nowadays, links in many networks are unstable that it will be changed with time or human will. As far as social network, links between one person and others will be weak and even disappear, due to mistrust, alienation, and sabotage. Likewise, in IERDCNs, connection probability exists in all collaboration. In this process, everyone likes a partner with powerful research capability, and never works together when either side of the R&D collaboration is failed. Additionally, enterprises have some interest in collaborating with institutes for technology R&D. As Table 4 and all parameters shows in previous part of this section, which means that there is no giant component and connection probability can make the component size smaller. Furthermore, self-governing choice of enterprises and institutes in practice reduces the chance that lots of enterprises collaborate with the same institute, which is relatively more conducive to technology R&D innovation of the whole networks. So by using connection probability, it could explain the process of autonomous choice and better describe the complicated network environment in the real world.

**Table 5**
Analytical and simulation results of IERDCN 1 with no giant component.

| | $\omega_1$=0.1, $\omega_2$=0.3 | | | | $\omega_1$=0.1, $\omega_2$=0.4 | | | |
|---|---|---|---|---|---|---|---|---|
| | $H_1^{'1}(1,1)$ | $H_1^{'2}(1,1)$ | $H_2^{'1}(1,1)$ | $H_2^{'2}(1,1)$ | $H_1^{'1}(1,1)$ | $H_1^{'2}(1,1)$ | $H_2^{'1}(1,1)$ | $H_2^{'2}(1,1)$ |
| Analytical results ($S_1$) | 1.6498 | 0.3517 | 2.8465 | 1.0124 | 5.1414 | 0.4690 | 13.6074 | 1.0220 |
| Simulation results($S_1$) | 1.6405 | 0.3492 | 2.8371 | 1.0000 | 5.1365 | 0.4635 | 13.5233 | 1.0000 |
| | $\omega_1$=0.2, $\omega_2$=0.3 | | | | $\omega_1$=0.2, $\omega_2$=0.4 | | | |
| | $H_1^{'1}(1,1)$ | $H_1^{'2}(1,1)$ | $H_2^{'1}(1,1)$ | $H_2^{'2}(1,1)$ | $H_1^{'1}(1,1)$ | $H_1^{'2}(1,1)$ | $H_2^{'1}(1,1)$ | $H_2^{'2}(1,1)$ |
| Analytical results ($S_1$) | 1.6579 | 0.9241 | 2.8661 | 1.0325 | 5.3193 | 1.2323 | 14.1124 | 1.0577 |
| Simulation results($S_1$) | 1.6463 | 0.9200 | 2.8401 | 1.0167 | 5.3115 | 1.2200 | 14.0067 | 1.0167 |

**Table 6**
Analytical and simulation results of IERDCN 1 with a giant component.

|  | $\omega_1=0.4, \omega_2=0.5$ | | $\omega_1=0.5, \omega_2=0.6$ | | $\omega_1=0.6, \omega_2=0.7$ | |
| --- | --- | --- | --- | --- | --- | --- |
|  | $S_1$ | $S_2$ | $S_1$ | $S_2$ | $S_1$ | $S_2$ |
| Analytical results ($S_1$) | 0.1522 | 0.2302 | 0.3214 | 0.4565 | 0.4633 | 0.6193 |
| Simulation results($S_1'$) | 0.1518 | 0.2300 | 0.3224 | 0.4650 | 0.4656 | 0.6350 |
|  | $\omega_1=0.7$ | $\omega_2=0.8$ | $\omega_1=0.8$ | $\omega_2=0.9$ | $\omega_1=0.9$ | $\omega_2=1.0$ |
|  | $S_1$ | $S_2$ | $S_1$ | $S_2$ | $S_1$ | $S_2$ |
| Analytical results ($S_1$) | 0.5905 | 0.7471 | 0.7074 | 0.8554 | 0.8156 | 0.9541 |
| Simulation results($S_1'$) | 0.5940 | 0.7800 | 0.7112 | 0.8950 | 0.8198 | 1.0000 |

Besides, connection probability could explain the disposable link, where if one link used it could not work again. Such as in IERDCNs, any side of networks could establish a partnership with each other, but the collaboration is not successful when without outputting patent or new technique. Some scholar as Fu et al. [16] use directed interacting networks for discussing this issue. However, unique contribution of our mathematical framework and arithmetic is providing another method to solve the situation that there is an outgoing edge with no return or opposite. Connection probability may be seen as precondition, whether R&D collaboration can generate new technique. And its value could refer to average success rate of R&D collaboration with knowledge diffusion. Furthermore, it also could give some reference basis for government in working out innovation policies. To percolating and promoting R&D collaboration for outputting as many new technologies and patent as possible, the government could moderately increase or reduce the connection probability. It is obvious that the collaboration between enterprises and institutes is important to technology R&D because of complementarily advantages. So we always have $\omega_2>\omega_1$ in IERDCNs as mentioned. After the results listed in Table 4, we give different analytical and simulation results (50 times) under four couple connection probability (see Table 5 and Table 6). Neither Table 4 nor Table 5 has a giant component. However, Table 6 shows the network have already undergone the phase transition. Comparing with Table 5, we find the relative growth rate of all average component sizes under $\omega_2$ is larger than under $\omega_1$. Furthermore, obviously, the $H_1^{'1}(1,1)$ and $H_2^{'1}(1,1)$ separately increase three nodes and 11 nodes when changing $\omega_2$ from 0.3 to 0.4, while remain the same when changing $\omega_1$ from 0.1 to 0.2. Similarly, the discrepancy between analytical results and simulation results enlarges with the increase of connection probability. And the precisions change respectively from 0.26% to 0.5% and 0.09% to 0.48%. All results are given to show the effectiveness and applicability of the proposed arithmetic.

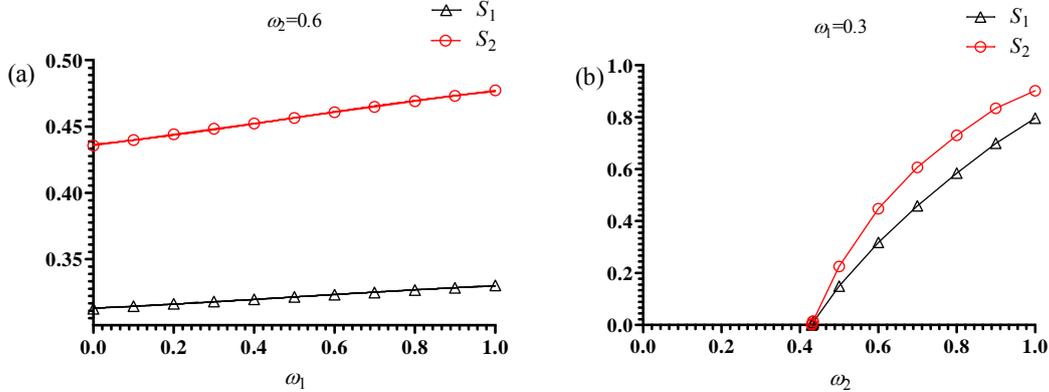

**Fig. 4.** Analytical results and simulation results of $S$ versus different $\omega_1$ and $\omega_2$, where $S$ denotes the probability that a randomly chosen node is a part of the giant component. (a) shows the probability when $\omega_2=0.6$ and $\omega_1$ change from 0 to 1. (b) shows the probability when $\omega_1=0.3$ and $\omega_2$ change from 0.4301 to 1.

Finally, using our arithmetic is easy to calculate every percolation threshold under some fixed connection probability. As Fig. 4 shows, we give two groups values to discuss how connecting probability of different networks affects whole network percolation. The threshold value (0.43) of IERDCN 1 is easy to be calculated according to Eq. (26) we discussed. Fig. 4(a) indicates that the probability that a randomly chosen a node belongs to the giant component is hardly influenced by $\omega_1$. Instead, Fig. 4(b) indicates that $\omega_2$ is the main cause. When let $\omega_2$ equal some fixed value and $\omega_1$ runs from 0 to 1, $S_1$ and $S_2$ are measured tiny change. However, if let $\omega_1$ equal some fixed value and $\omega_2$ runs from 0.4301 to 1, we can see double inward curves and curvature of $S_2$ changes greatly. That's why we believe that the inter-firm competition and desire for new technique makes collaborations between institutes and enterprise are very important to IERDCNs. And IERDCN 2 is considered as an ideal state for both sides. In other words, every enterprise can find one or two partners from institutes, while institutes have equal opportunities to collaborate with enterprises. There two phrases may be experienced such as IERDCN 1 and 2. In the first one, both sides strongly incline to collaborate with famous and technology strength partner, which is very much like celebrity effect. In another, technological monopoly advantages difficult to maintain with knowledge and technique of the industry flow in networks, especially institutes have accumulated a wealth of R&D experience, mastered the advanced technology in this field after a period of collaboration. And some enterprises and institutes have established long-term R&D collaboration relationship, which helps achieve the ideal state. We find that the lower the interdependence between enterprises or institutes, the higher percolation on IERDCNs. That's why the networks have a giant component with lesser inner connection probability and taller interactive one. And it is common in many high-tech industries, such as IERDCN 1 only 30 links between enterprises and institutes is enough to maintain the supercritical regime.

## 5. Conclusion

The connection probability considered a couple of nodes with temporary edge but cannot be effective connect to each other. It is so important to a dynamic evolution network of R&D collaboration, which is a knowledge dissemination network. From the perspective of the success or failure of the collaboration, it illustrates knowledge and techniques cannot be able to disseminate with invalid collaboration.

Our mathematical framework and arithmetic discussed in this paper are near-perfect to the real state of IERDCNs, which are accurate as the results mentioned in Section 4 shows. In this case, despite the sizes of IERDCNs are not very large and exist discrepancies, it does not greatly damage the validity of arithmetic. So we may shelve generating larger networks by the simulation for a while. However, we have simulated 50 times of them randomly to study, sometimes the discrepancies are artificially enlarged. We believe the discrepancies will be likely to be quite minimal if all the samples are considered.

In this study, we have investigated two types of mathematical framework and arithmetic based on generating functions for analyzing IERDCNs. Furthermore, we have given the necessary and sufficient conditions for the absence of the system-wide giant components. Using our arithmetic can calculate the corresponding parameters in the sub-critical and supercritical regimes. Through two actual IERDCNs, application and validity of our mathematical framework and arithmetic were discussed. It is quite clear that interactive connection probability of networks is a determinant factor of the percolation, while inner connection probability has less influence. Some reasonable and helpful advices are given to promote regional technological R&D and innovation. By adjusting the probability values, we also found the supercritical regime of the whole network is maintained mainly collaborated between enterprises and institutes.

## Acknowledgments

This work was partly supported by the National Natural Science Foundation of China under Grant No. 70972115.